\documentclass[useAMS,usenatbib]{mn2e}
\usepackage{deluxetable}
\usepackage{natbib}
\usepackage{graphics}
\usepackage{graphicx}
\usepackage{amsbsy}
\usepackage{amssymb}
\usepackage{subfigure}
\usepackage{color}

\newcommand{\be}{\begin{equation}}
\newcommand{\ee}{\end{equation}}

\title{Short gamma ray bursts: formation and offsets}
\author[C. Boylan, Y. Li, X.L. Fan, I.S. Heng]{C. Boylan, Y. Li, X.L. Fan\thanks{E-mail:
xilong.fan@glasgow.ac.uk}, I.S. Heng \\
SUPA, School of Physics and Astronomy, University of Glasgow, Glasgow G12 8QQ, United Kingdom}
\begin{document}

\date{Received xxxx / Accepted xxxx}

\pagerange{\pageref{firstpage}--\pageref{lastpage}} \pubyear{2014}

\maketitle

\label{firstpage}

\begin{abstract}
Short gamma ray bursts have been observed a variety of galaxies types with varying angular offsets from the centre of their host galaxies.
To investigate the properties of short gamma ray burst offsets, a sample of short gamma ray bursts with host galaxies has been gathered.
Two formation channels proposed to explain the observed offsets of short gamma ray bursts from their host galaxies are discussed.
The classification of short gamma ray bursts into these formation channels is demonstrated for short gamma ray bursts with host galaxies.
The possibility of faint dwarf galaxies as host environments for the observed short gamma ray bursts is also investigated.
However, by extrapolating the distribution of dwarf galaxies around the Milky Way to redshifts corresponding to the short gamma ray bursts, we show that it is unlikely that dwarf galaxies are the hosts of short gamma ray bursts.

\end{abstract}


\section{Introduction}

Gamma Ray Bursts (GRBs) are non-repeating, highly energetic events that emit mainly gamma rays over a short timeframe. 
First detected by the Vela satellites in 1969 \citep{Klebesadel1973ApJ...182L..85K}, hundreds of theoretical models were emerged to explain these bursts while little data was gained to verify these assumptions. However, in 1997, the first optical and X-ray afterglows were detected with direct measurements of their redshifts through optical spectroscopy which leads to an intense investigation of GRBsÕ progenitor channels, energy source and origin systems. 

GRBs can be classified into two categories, mainly based on duration and spectral hardness: the short-hard and long-soft bursts with a separation at about 2 sec \citep{Kouveliotou1993ApJ...413L.101K}.
The longer GRBs (LGRBs) can extend to hundreds of seconds. They originate from the brightest regions of galaxies and closely track stellar mass \citep{Wainwright2007ApJ...657..367W}. For LGRBs that originate from distances close enough for supernova detection to be possible it has been observed that the events can usually be associated with a Type 1b or Type 1c supernova. The almost unanimous consensus among the astrophysics community is that the collapse of rapidly-rotating massive stars or {\it collapsars} are LGRB progenitors~\citep{MacFadyen1999ApJ...524..262M,Fruchter2006Natur.441..463F}. Short gamma ray bursts (SGRBs), on the other hand, can last as short as ~ milliseconds and have harder spectra and lower fluences than their longer duration counterparts. SGRB progenitors are thought to be the coalescence of compact object binaries (with neutron star and/ or black hole constituents Ð DNS/NS-BH) in the case of SGRBs \citep{Paczynski1991AcA....41..257P}

SGRBs are significantly fainter than their long duration counterparts, and were, hence, observed at a relatively low event rate with large and uncertainties in source sky location~\citep{Hurley2002ApJ...567..447H}. The launch of NASAÕs Swift satellite in late 2004, which provided the first chance for detecting accurate positions rapidly for SGRBs, indeed led to the detection of the first X-ray, optical and radio afterglows in the following years \citep{Berger2005Natur.438..988B}. As an increasing number of SGRBs are localized with measurements of redshift and host galaxy data, they have been in both star-forming and elliptical (old stellar population) galaxies and not just in the galactic centre but sometimes in the outer regions of the galaxy and in some cases even in the galactic halo \citep{Belczynski2006ApJ...648.1110B,Berger2011NewAR..55....1B,Fong2013ApJ...769...56F}.  Up to now, only one  SGRB  (GRB130603B) has been observed to be associated with a `kilonova' \citep{Tanvir2013Natur.500..547T}. The observation that would furthermore confirm that the coalescence of compact binaries is the correct progenitor system would be the coincident detection of gravitational waves from the event. However, this confirmation is not expected to be made in the next few years so at present the best method for deducing the most likely progenitor system is through statistical analysis of the locations of the SGRBs. 

In this paper, we discuss the SGRB formation scenarios proposed to explain the observed locations with respect to their host galaxies. In particular, two scenarios, primodial and dynamical, for explaining the observed SGRB {\it offsets} are explored and applied to a subset of SGRBs with host galaxies. We then investigated the possibility that SGRBs are hosted by faint dwarf galaxies around its brighter host galaxy before discussing our observations.


\section{data set}
To investgate  the formation of SGRBs by their offset properties based on Table \ref{yl_result} and reduce the uncertity, we select the GRBs from the data set  shown in \cite{Fong2013ApJ...769...56F}.  The sample of 7  SGRBs,  which do not have  extended emissions, are locating by more than one camera  and their hosts's redshift and morphology are avaliable,   are shown in  Table \ref{YL_grb}.

As discussed in the previous section, short gamma ray bursts are not often observed to lie within the centre of their host galaxy but rather towards the outer regions or outside of the galaxy entirely. One aim of this research was to determine whether the short GRB progenitors could be located in satellite galaxies in orbits around their host galaxies. In order to carry out this investigation, data from several papers were amalgamated to construct a dataset which comprised of short GRB observations which were accurately associated with a host galaxy. Each event was observed with an angular offset from the host galactic centre. The events and their data are shown in Table \ref{grb_cb} \citep{Fong2010ApJ...708....9F,Fong2013ApJ...776...18F}. It can be seen that most of the events that have been successfully associated with a host galaxy are from low redshifts (z $<$ 1). In order to determine whether or not these observed angular offsets could be indicative of a satellite/dwarf galaxy host a typical satellite galaxy distribution (of distances from galactic centre) must be obtained for comparison.

\section{GRB formation and offsets}
These are the two main channels for compact binariesÕ formations in different scenario:
through the evolution of massive stars in primordial binaries rising in the galactic field \citep{Narayan1992ApJ...395L..83N}, and from three (or even more)-body dynamical interactions among stars and compact remnants in globular clusters (GCs)\citep{Grindlay2006NatPh...2..116G}. To reproduce the redshift distribution data of SGRBs from Swift, both formation channels are important \cite{Salvaterra2008MNRAS.388L...6S}.

\subsection{ Kicked  or  born? }
A binary star systems behaviour and evolution are determined by the initial masses of the stars, their semi-major axis distance and the eccentricity of the binary at the time when the system is formed. In the first stage of the binaryÕs lifetime there is a small amount of mass transfer between the two stars which acts to circularize the orbit of the binary and also cause the stars to spiral towards each other \citep{Fong2010ApJ...708....9F}.
Then, when one of the stars in the system reaches supernova the resulting NS often receives an asymmetric impulse as a result of the event. The Òkick velocityÓ which a NS receives can vary significantly and it has been shown that a Maxwellian velocity distribution fits well with observations \citep{Bloom1999MNRAS.305..763B}. Complex modelling has shown that the semi-major axis of the orbit is altered and that the binary system as a whole obtains a kick velocity in a random direction with a median velocity of around 300 km s-1. These models consider the birth-rate of binary systems, initial system velocity, initial position in the galaxy, lifetime of the stars and the merger rate of the neutron stars which is an average of one hundred million years. The results predict that the median radial distance from the galactic centre for a NS-NS merger is 10kpc and that ninety percent of the mergers should occur within 30kpc of the host (this corresponds to 4 arc seconds at z = 1).These results match well with the sample of GRBs used in this research and act to confirm the conclusion that the short gamma ray bursts are not situated in dwarf galaxies. The typical dwarf galaxy distribution has ninety five percent of the satellites at a distance greater than 20kpc from the galactic centre.
\begin{table*}
\scriptsize
\caption{Complete SGRB Host Galaxy Sample}\label{YL_grb} 
\begin{tabular}{l|lll|lll|ll}
\hline
\hline
GRB&	T$_{90}$ &host Type$^{a}$	&z&	offset\\
&	(s) &	 &  &(kpc)\\
\hline
051221A	&1.4&	L&	0.546&	0.76\\
070429B	&0.5&	L	&0.9023	&4.7\\
070724A	&0.4&	L	&0.457	&4.76\\
071227	&1.8	&L&	0.381	&15.0\\
090426A	&1.3&	L&	2.609&	0.8\\
100117A	&0.3	&S	&0.915	&0.5\\
111117A	&0.5	&L	&1.3	&10.5\\
 \hline
\end{tabular}\\
$^{a}$ S(L) refers to the small (large) galaxy model \citep{Belczynski2006ApJ...648.1110B}. Data are selected from \cite{Fong2013ApJ...769...56F} (see text for detail).
\end{table*}
\begin{table*}
\scriptsize
\caption{ SGRB offset and redshift }\label{grb_cb} 
\begin{tabular}{l|lll|lll|ll}
\hline
\hline
GRB &	Offset (arcesecs) &Redshift\\
\hline
061201 &	16.25	&0.111\\
070429B&	1.46	        &0.902\\
070714B	&1.55	&0.923\\
070724A	&0.94	&0.4571\\
070809	&5.63	&0.473\\
071227	&2.98	&0.381\\
080905A	&8.29	&0.1218\\
090510	&1.33	&0.903\\
090515	&13.98	&0.403\\
100117A	&0.17	&0.915\\
130603B	&1.05	&0.3564\\
070707	&0.4	         &3.6\\
090305A	&0.43	&4.1\\
090426	&0.06	&2.609\\
 \hline
\end{tabular}\\
Each event was observed with an angular offset from the host galactic centre.  Data are are from  \cite{Fong2010ApJ...708....9F,Fong2013ApJ...776...18F}.
\end{table*}

Population synthesis models has been made to computed for theoretical spatial distribution of primordial SGRBs generated from isolated galaxies of different types and sizes \citep{Belczynski2006ApJ...648.1110B}. Considering three different galaxy types (elliptical, spiral and starburst) and two kinds of sizes (small hosts with viral masses of $\sim10^9M_{\odot}$ and large ones of $\sim10^{12}M_{\odot}$, respectively), two windows for offsets (first from 0 to 10 kpc, and the second between 10 and 100 kpc) are selected thus SGRBs can be located well inside their host galaxies. However, the "kick velocity" explanation  for  offset of GRBs are inconsistent with observations of double neutron star systems in the Milky Way. The Hulse-Taylor binary pulsar,  together with  other known double neutron star systems in the Milky Way appear to be low kick-velocity  ($\leq 50$ km s$^{-1}$) system.  Thus, double neutron star systems are expected to remain within the central region of their parent (spiral or star-forming) galaxies \citep{Grindlay2006NatPh...2..116G}. 

Regarding of dynamical origin of SGRBs: they came from double neutron stars and went through dynamical channels (usually as a result of few body interactions). While binaries to be released with large recoil velocities are almost impossible, underlying globular clusters (GCs) spatial distribution should be taken into consideration.
In the case of host galaxies in clusters, GC distribution is cut by tidal truncation within the galaxies, which leaves the offset in 10-100 kpc an unclear window. On the one hand, only 10-20\% of massive ones are in this offset window; on the other hand, the bulk of GCs from small galaxies are less than 10 kpc. Bound GCs mostly extend up to 50-100 kpc both in large and small galaxies \citep{Salvaterra2010MNRAS.406.1248S}. Speaking of the SGRBs from isolated galaxies, as both theoretical and observational support the existence of numerous intra-cluster GCs (ICGCs), GCs are believed to have a wider distribution so as to explain large potential offsets beyond the theory of natal kicks \citep{Salvaterra2010MNRAS.406.1248S}. From theoretical estimation ICGCs occupy $\sim$ 30\% despite of the cluster total mass of the GCsÕ spatial distribution. Since observations have provided an indication of ICGCsÕ existence, they should be found far from center of host cluster and spread throughout the cluster volume as expected.  The main results of all above  studies by far are listed in Table. \ref{yl_result}  \citep{Salvaterra2010MNRAS.406.1248S}. We will compare the samples shown in Table \ref{YL_grb} with results of  \cite{Salvaterra2010MNRAS.406.1248S}.

As shown in  Table \ref{YL_grb}, In the cases of GRB 051221A, GRB 070429B, GRB 070724A and GRB 090426A, which are all discovered within Òlarge-modelÓ host galaxies and their offsets from each hosts fall in the scope of 0-10 kpc, these SGRBs are likely to result from the primordial channel. The indications of this kind of SGRBs are relatively typical, since the intrinsic absorption in gamma-ray spectrum measurements, the redshift data from observation and their positions of
optical afterglow (usually inside their host galaxies) are all symbolization of a primordial binary system.
For GRB 100117A, the relatively small offset ($\sim$0.5 kpc) away from its ÒsmallÓ host galaxy (which identified as elliptical type) only suggests the possibility of a dynamical origin can be 2.5 times than that from the primordial channel.
GRB 071227 and GRB 111117A are even more interesting, given that the classification of such kind of SGRBs is still controversial. Consider GRB 071227 which was already firmly classified as an SGRB \citep{Levan2008MNRAS.384..541L}, both the formation channels of primordial or dynamical conform to the situation: 15 kpc offset from a relatively large (r $\sim$15 kpc) spiral galaxy, making the classification of GRB 071227 and GRB 111117A not so straightforward. However, our best speculation for GRB 071227 based on its relative position from the host, together with the position of afterglow was observed within the hostÕs galactic plane, provides a possible explanation of a primordial origin.
From the samples above we have seen the analysis results a powerful tool to identify different cases of their origins of the offsets.

\subsection{Faint  Dwarf galaxies as the physical hosts  of offset SGRBs}

Why the progenitor kicks would possess such a broad distribution remains unclear. It is likely that the distribution is a function of the binary system and host galaxy size, rather than the progenitor model. SGRBs that have not been robustly associated to a host galaxy, due to their apparently large offsets, are attributed to the nearest galaxy of lowest probability of chance coincidence \citep{Bloom2002AJ....123.1111B}. However these SGRBs may, instead, have be very faint galaxies currently hidden from observation by an active interstellar foreground.
  
  \begin{table*}
\scriptsize
\caption{Percentages of SGRBs with offset in the three domains for the two formation channels and the two galaxy models.}\label{yl_result}
\begin{tabular}{l|lll|lll|ll}
\hline
\hline
      &  \multicolumn{2}{c}{0-10kpc} &  \multicolumn{2}{c}{10-100kpc} &  \multicolumn{2}{c}{$>$100kpc}\\
    \cline{1-7}\\
   & small &large &  small & large & small & large\\
Primordial  &0.15  &0.60  &0.10 &  0.20&0.75&0.20   \\   
Dynamical &0.40 &0.05  &0.55 &  0.30&0.05&  0.65 \\      
 \hline
\end{tabular}\\
The percentages are computed for SGRBs resulting from primordial binaries and from the dynamical channel. Left panel is for isolated galaxies with small/large galaxy model. Note that the table does not provide the relative contribution from the two channels but only the distribution in the three offset intervals. Table are from \cite{Salvaterra2010MNRAS.406.1248S}
\end{table*}

 Dwarf galaxies are gravitationally bound systems (typically $\sim$1 billion stars) orbiting larger host galaxies. The satellite galaxy population of our own galaxy and that of M31, Andromeda are well known and the distances to each of the dwarf galaxies may be taken from many sources and are generally accurate to within around 5\%. Fig 1 shows the distribution of dwarf galaxies around the two galaxies \citep{Metz2007MNRAS.374.1125M}.

 The limitation  of  current observation $< 25.5$ AB mag \citep{Fong2010ApJ...708....9F}  shows another  channel of the nature of  GRB offsets:  their physical hosts are faint dwarf  galaxies.  In particular, the apparent magnitude of Andromeda itself is 3.44 and the average apparent magnitude of its dwarf galaxies is 12.45 Table\ref{m31_ab}. For an approximate analysis then it can be taken that an average dwarf galaxy should be nine orders of magnitude dimmer than its host as shown in column four of Table \ref{host_ab}. This suggests that the brightest host galaxy from this sample should have dwarf galaxies with an AB Mag of around twenty-five but the average host galaxy is expected to have dwarf galaxies with magnitudes on the order of thirty.

\begin{table*}
\scriptsize
\caption{Apparent magnitude data for Andromeda and it's satellite galaxies.}\label{m31_ab}
\begin{tabular}{l|lll|lll|ll}
\hline
\hline
 Galaxy &	Apparent Magnitude\\
\hline
Andromeda & 3.44\\
M32 &9.2\\
M110	&9.4\\
NGC 185	&11\\
NGC 147	&12\\
Andromeda I	&13.2\\
Andromeda II	&13\\
Andromeda III	&10.3\\
Andromeda V	&15.4\\
Andromeda VI	&14.5\\
Andromeda VIII &	9.1\\
Andromeda IX	&16.2\\
Andromeda X	&16.2\\
 \hline
\end{tabular}
\end{table*}

\begin{table*}
\scriptsize
\caption{AB Magnitude data for host galaxies of observed short GRBs.	}\label{host_ab}
\begin{tabular}{l|lll|lll|ll}
\hline
\hline
GRB&	Redshift	&Host AB &	Host AB\\
       &                      & Mag &   Mag+9\\
\hline
050509b  &0.226	&16.32	&25.32\\
050709&	0.1606	&21.09	&30.09\\
050724&	0.257	&19.98	&28.98\\
051210&	$>$1.4	&21.14	&30.14\\
051221a&	0.5465	&21.86	&30.86\\
060121	&?	&26.22&	35.22\\
060313	&?	&26.38	&35.38\\
060502b	&?	&17.88	&26.88\\
061006	&0.4377&	21.67	&30.67\\
061201	&0.111	&18.17	&27.17\\
 \hline
\end{tabular}
\end{table*}

\begin{figure*}
\begin{center}
  \includegraphics[width=\textwidth]{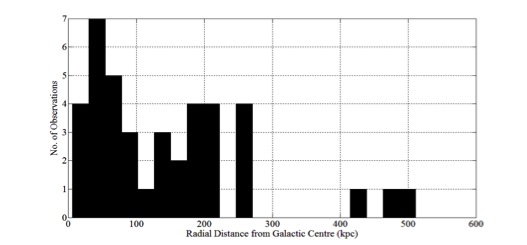}
  \caption{Histogram showing the distribution of dwarf galaxies around the Milky Way and Andromeda galaxies in kiloparsecs.\label{df_hist}}
\end{center}
 \end{figure*}

From Fig \ref{df_hist}, It can be seen from the distribution that of the 40 well documented dwarf galaxies of the two galaxies only two lie within 20kpc of the galactic centre ($95\% \geq 20$kpc) and that the most probable location for a dwarf galaxy is within the radial distance of 30-60kpc from the galactic centre. Fig \ref{df_hist} also tentatively reveals a Òcut-offÓ distance somewhere between 200kpc and 300kpc at which point the number of dwarf galaxies drops. This feature is due to the virial radii of the two galaxies, the Milky Way and Andromeda, which have virial radii of around 200kpc and 260kpc respectively. The virial radius is the orbital radius at which an object is in a stable orbit and therefore we would expect to see a sharp reduction in satellites beyond this distance. As the Milky Way and Andromeda are fairly typical galaxies and their satellite distributions are fairly similar it is reasonable to assume that this distribution is representative of a typical satellite galaxy distribution. This assumption allows a Òdwarf galaxy windowÓ to be defined. For any observed galaxy it is likely that its satellite galaxies will lie between 20kpc and 200kpc from the galactic.
 
 By using the angular size - redshift relation,  the ``dwarf galaxy window" can be converted from a radial distance window to an angular offset window as a function of redshift. By superimposing the observed short gamma ray bursts onto the angular offset window it is possible to see whether or not the GRBs are observed at separations coincident with where dwarf galaxies would be expected.
 
\begin{figure*}
\begin{center}
  \includegraphics[width=\textwidth]{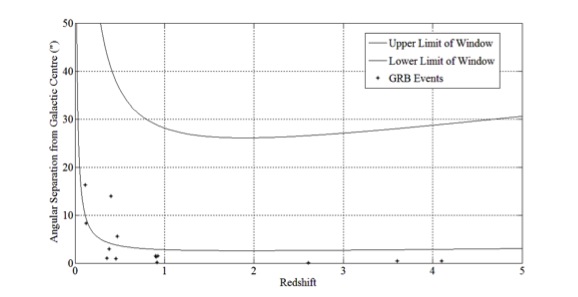}
  \caption{The star markers represent the observed GRB angular offsets and the two lines represent the expected minimum and maximum range of dwarf galaxy offsets.\label{angle_1}}
\end{center}
\end{figure*}

\begin{figure*}
\begin{center}
  \includegraphics[width=\textwidth]{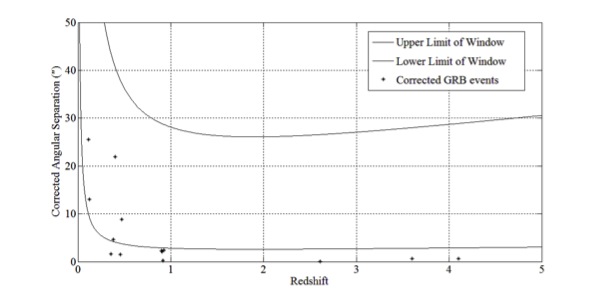}
  \caption{ The star markers represent the observed GRB angular offsets with the 3D correction applied and the two lines represent the expected minimum and maximum range of dwarf galaxy offsets..\label{angle_2}}
\end{center}
\end{figure*}

Fig \ref{angle_1} shows that only three of the fourteen observations have offsets greater than the lower boundary of the dwarf galaxy window. Given that ninety five percent of dwarf galaxies lie beyond this lower limit, it seems that it is unlikely that these short GRBs are indeed originating from satellite galaxies. However, there is one more aspect of the observations to be considered. The geometry of the observer, galactic centre and observed event will determine what fraction of the actual separation is observed. A telescope will only observe the projection of the angular offset which is perpendicular to the line of sight.
If the GRBs occur at a radius, $R$, from the galactic centre and at an angle, $\theta$ , from the plane perpendicular to the line of sight (the angle, $\phi$, around the plane does not affect the magnitude of the offset measured) then the observed angular offset will only correspond to $ R \cos(\phi)$.
Although it is impossible to determine $\theta$ for any single observation, if one considers the average projection for all values of $\theta$ and then uses this to scale up each of the observed angular offsets then a more physical distribution can be obtained.
On each side of the plane perpendicular to the line of sight, $\theta$ can be any value between $0$ and $\pi$. Thus, the average fraction of the separation that is observed is given by: 
\be
\frac{1}{\pi} \int_0^{\pi} | \cos(\theta)| \, d\theta =\frac{2}{\pi}
\ee 
Therefore if the GRB angular separations plotted in Fig \ref{angle_1} are multiplied by a factor of  $\frac{\pi}{2}$ then a better comparison can be made. 
Fig \ref{angle_2} shows that now five of the fourteen short GRB observations lie within the dwarf galaxy window. This figure is still substantially different from what would be expected if the progenitors of these short GRBs were originating from dwarf galaxies. Even from this small statistical analysis it is clear that the observations have a tendency to lie nearer to their host galaxies than the typical satellite galaxy population. Although more data is needed to confirm this conclusion, this research may tentatively conclude that the observed angular offset distributions of short gamma ray bursts suggest that the events are not occurring in dwarf galaxies orbiting the associated host galaxy.

 \section{Conculation and discusstion}
 By comparison with a typical satellite galaxy population this statistical analysis of the angular offsets of fourteen  SGRBs has revealed that the observed separations between the events and the galactic centre are not explained by the fact that the SGRBs are occurring in dwarf galaxies. This study has shown that less than half of the SGRBs lie beyond the offset that we would expect ninety-five percent of dwarf galaxies to lie beyond.   Although this study appears to have answered the question of whether or not short gamma ray bursts could originate from dwarf galaxies, the accuracy of the analysis leaves room for future improvement. In the next few years, it is expected that many more  SGRBs will be localized to host galaxies and it is not unrealistic to expect that there will be data for hundreds of events by the end of the decade. This will allow for a much more detailed statistical analysis and the demographics of the events will become more readily apparent.
Aside from the limited SGRB data, this study could be improved by developing a model for satellite galaxy distributions from a more diverse range of galaxy sizes rather than assuming that all galaxies will have similar dwarf galaxy populations to Andromeda and the Milky Way. It would be expected that smaller, less luminous galaxies would have satellite galaxies which lie at shorter radial distances than larger galaxies would. Thus, if the absolute magnitude of the host galaxies can be calculated from the observations then an estimate can be made of their size and a corresponding dwarf galaxy distribution could be applied. This method would allow more confident conclusions to be drawn from the research.

By applying the limit samples of seven  SGRB offsets and their  hosts to the compact binaries formation channels, four  and one samples are prefer  the primordial channel  and dynamical  channel of compact binaries formation scenario, respectively. Two sample  can not distringush the formation   channel. We hope more date in the future will give more constraints. 

\bibliographystyle{apj}
\bibliography{ref}
\label{lastpage}

\end{document}